\documentclass[12pt]{article}

\usepackage{graphicx}

\def\Color#1#2{#2}

\def\DR#1{\Color{0.27 1.0 1.0 0.32} {#1}}

\def\DB#1{\Color{1.0 0.7 0.0 0.6} {#1}}
\def\DG#1{\Color{0.75 0. 0.75 0.75} {#1}}

\usepackage{scicite}

\usepackage{times}

\newenvironment{sciabstract}{%
\begin{quote} \bf}
{\end{quote}}

\newcounter{lastnote}
\newenvironment{scilastnote}{%
\setcounter{lastnote}{\value{enumiv}}%
\addtocounter{lastnote}{+1}%
\begin{list}%
{\arabic{lastnote}.}
{\setlength{\leftmargin}{.22in}}
{\setlength{\labelsep}{.5em}}}
{\end{list}}

\title{Cosmic Mnemonics}

\author
{Douglas Scott,$^{1\ast}$ Ali Narimani,$^{1}$ Don N. Page$^{2}$\\
\\
\normalsize{$^{1}$Department of Physics \& Astronomy, University of British
Columbia,}\\
\normalsize{Vancouver, BC, Canada\ V6T 1Z1}\\
\normalsize{$^{2}$Department of Physics, University of Alberta,}\\
\normalsize{Edmonton, AB, Canada\ T6G 2E1}\\
\\
\normalsize{$^\ast$To whom correspondence should be addressed; E-mail:
 dscott@phas.ubc.ca}
}

\date{}

\def\dddot#1{
  \def\ddd{.\mkern-1.8mu.\mkern-1.8mu.\mkern-1mu}
\mathchoice
  {\vbox{\offinterlineskip
  \halign{\hfil##\hfil\crcr
  $\smash{\displaystyle\ddd}$\crcr
  \noalign{\vskip0.4ex}
  $\displaystyle{#1}$\crcr}}}
  {\vbox{\offinterlineskip
  \halign{\hfil##\hfil\crcr
  $\smash{\textstyle\ddd}$\crcr
  \noalign{\vskip0.38ex}
  $\textstyle{#1}$\crcr}}}
  {\vbox{\offinterlineskip
   \halign{\hfil##\hfil\crcr
  $\smash{\scriptstyle\ddd}$\crcr
  \noalign{\vskip0.3ex}
  $\scriptstyle{#1}$\crcr}}}
  {\vbox{\offinterlineskip
  \halign{\hfil##\hfil\crcr
  $\smash{\scriptscriptstyle\ddd}$\crcr
  \noalign{\vskip0.17ex}
  $\scriptscriptstyle{#1}$\crcr}}}}
\def\ddddot#1{
  \def\dddd{.\mkern-1.6mu.\mkern-1.6mu.\mkern-1.6mu.\mkern-2mu}
\mathchoice
  {\vbox{\offinterlineskip
  \halign{\hfil##\hfil\crcr
  $\smash{\displaystyle\dddd}$\crcr
  \noalign{\vskip0.4ex}%
  $\displaystyle{#1}$\crcr}}}
  {\vbox{\offinterlineskip
  \halign{\hfil##\hfil\crcr
  $\smash{\textstyle\dddd}$\crcr
  \noalign{\vskip0.38ex}
  $\textstyle{#1}$\crcr}}}
  {\vbox{\offinterlineskip
  \halign{\hfil##\hfil\crcr
  $\smash{\scriptstyle\dddd}$\crcr
  \noalign{\vskip0.3ex}
  $\scriptstyle{#1}$\crcr}}}
  {\vbox{\offinterlineskip
  \halign{\hfil##\hfil\crcr
  $\smash{\scriptscriptstyle\dddd}$\crcr
  \noalign{\vskip0.17ex}
  $\scriptscriptstyle{#1}$\crcr}}}}
\def\dddddot#1{
  \def\ddddd{.\mkern-1.4mu.\mkern-1.4mu.\mkern-1.4mu.\mkern-1.4mu.\mkern-3mu}
\mathchoice
  {\vbox{\offinterlineskip
  \halign{\hfil##\hfil\crcr
  $\smash{\displaystyle\ddddd}$\crcr
  \noalign{\vskip0.4ex}%
  $\displaystyle{#1}$\crcr}}}
  {\vbox{\offinterlineskip
  \halign{\hfil##\hfil\crcr
  $\smash{\textstyle\ddddd}$\crcr
  \noalign{\vskip0.38ex}
  $\textstyle{#1}$\crcr}}}
  {\vbox{\offinterlineskip
  \halign{\hfil##\hfil\crcr
  $\smash{\scriptstyle\ddddd}$\crcr
  \noalign{\vskip0.3ex}
  $\scriptstyle{#1}$\crcr}}}
  {\vbox{\offinterlineskip
  \halign{\hfil##\hfil\crcr
  $\smash{\scriptscriptstyle\ddddd}$\crcr
  \noalign{\vskip0.17ex}
  $\scriptscriptstyle{#1}$\crcr}}}}
\def\ddddddot#1{
 \def\dddddd{.\mkern-1.4mu.\mkern-1.4mu.\mkern-1.4mu.\mkern-1.4mu.\mkern-1.4mu.\mkern-4mu}
\mathchoice
  {\vbox{\offinterlineskip
  \halign{\hfil##\hfil\crcr
  $\smash{\displaystyle\dddddd}$\crcr
  \noalign{\vskip0.4ex}%
  $\displaystyle{#1}$\crcr}}}
  {\vbox{\offinterlineskip
  \halign{\hfil##\hfil\crcr
  $\smash{\textstyle\dddddd}$\crcr
  \noalign{\vskip0.38ex}
  $\textstyle{#1}$\crcr}}}
  {\vbox{\offinterlineskip
  \halign{\hfil##\hfil\crcr
  $\smash{\scriptstyle\dddddd}$\crcr
  \noalign{\vskip0.3ex}
  $\scriptstyle{#1}$\crcr}}}
  {\vbox{\offinterlineskip
  \halign{\hfil##\hfil\crcr
  $\smash{\scriptscriptstyle\dddddd}$\crcr
  \noalign{\vskip0.17ex}
  $\scriptscriptstyle{#1}$\crcr}}}}

\newbox\tablebox    \newdimen\tablewidth
\def\leaderfil{\leaders\hbox to 5pt{\hss.\hss}\hfil}
\def\endtable{\tablewidth=\hsize
    $$\hss\copy\tablebox\hss$$
    \vskip-\lastskip\vskip -2pt}

\def\tablenote#1 #2\par{\begingroup \parindent=0.8em
    \abovedisplayshortskip=0pt\belowdisplayshortskip=0pt
    \noindent
    $$\hss\vbox{\hsize\tablewidth \hangindent=\parindent \hangafter=1 \noindent
    \hbox to \parindent{\sup{\rm #1}\hss}\strut#2\strut\par}\hss$$
    \endgroup}
\def\doubleline{\vskip 3pt\hrule \vskip 1.5pt \hrule \vskip 5pt}

\begin{document} 


\baselineskip24pt


\maketitle


\begin{sciabstract}
Our current description of the large-scale Universe is now known with
a precision undreamt of a generation ago.  Within the
simple standard cosmological model only six basic parameters are required.
The usual parameter set includes quantities most directly probed
by the cosmic microwave background, but the nature of these quantities is
somewhat esoteric.
However, many more numbers can be derived that quantify various aspects of
our Universe.  Using constraints from the \textit{Planck\/} satellite, in
combination with other data sets, we explore several such quantities,
highlighting some specific examples.
\end{sciabstract}

\section*{Introduction}

Astrophysicists are currently drowning in unprecedented amounts of data,
including some that can be used to pin down the parameters
describing the statistical properties of the entire large-scale Universe
within the context of a simple model.  As a
result of these data, many scientists are hailing this
as the `era of precision cosmology' \cite{TurnerTyson}.

This precision has taken another step forward with the recent publication of
cosmological results from the cosmic microwave background (CMB)
satellite {\it Planck\/}
\cite{PlanckMission,PlanckParameters}.  The {\it Planck\/}
findings further underscore our rather full
accounting of the cosmic energy budget and an assessment of
how fast the Universe is
expanding, as well as other quantities describing the density perturbations
laid down at early times that grew into today's
astronomical structures.

An often-stated result, forming the focus of the main
cosmological parameter paper from {\it Planck}, is that {\it merely six\/}
numbers are sufficient to parameterise the `Standard Model of Cosmology'
(SMC, see Refs.~\cite{Triangle,SMC} and reviews in Ref.~\cite{RPP}).  However,
the significance of this tour de force of modern physics is undermined
by the difficulty of describing these parameters to a non-specialist --
the Universe on the largest scales is fully characterised
by the values of $\Omega_{\rm b}h^2$, $\Omega_{\rm c}h^2$, $\theta_\ast$,
$A_{\rm s}$, $n$ and $\tau$ (presented in Table~1 and
described below), all of which need considerable explanation.  Moreover,
the statement that the set contains only six parameters is a little
misleading for several reasons.  First of all, there are {\it other\/}
parameters that are fixed to their default values within the SMC.  These
include the overall curvature of space, the required mass of additional
species such as neutrinos, whether the dark energy density evolves,
and the existence of other types of fluctuations
in the early Universe.  Secondly, several parameters
are determined by astrophysical measurements {\it other\/} than CMB temperature
anisotropies.  These include the overall temperature of the CMB today,
the abundance of light elements such as helium, and the
numbers that describe the whole of the rest of physics!  And thirdly,
although six parameters may be sufficient within the SMC, the choice of
which parameters to include in that set is not unique.  Plenty of
interesting numbers can be {\it derived\/} from those most naturally
measured quantities.  A good example is the age of the Universe,
which is not directly determined from CMB measurements but is easy to
calculate once the SMC parameters have been pinned down.

\medskip
\nointerlineskip
\setbox\tablebox=\vbox{
\newdimen\digitwidth 
\setbox0=\hbox{\rm 0} 
\digitwidth=\wd0 
\catcode`*=\active 
\def*{\kern\digitwidth} 
\newdimen\signwidth 
\setbox0=\hbox{{\rm +}} 
\signwidth=\wd0 
\catcode`!=\active 
\def!{\kern\signwidth} 
\newdimen\pointwidth 
\setbox0=\hbox{\rm .} 
\pointwidth=\wd0 
\catcode`?=\active 
\def?{\kern\pointwidth} 
\halign{\tabskip=0.0em\DB{#}\hfil\tabskip=1.0em&
\DR{#}\hfil\tabskip=1.0em&
\DG{#}\hfil\tabskip=0.0em\cr
\noalign{\doubleline}
\multispan3\hfil Standard cosmological parameters\hfil\cr
\noalign{\vskip 3pt\hrule\vskip 4pt}
Parameter& Description& Value\cr
\noalign{\vskip 3pt\hrule\vskip 4pt}
$\Omega_{\rm b}h^2$& Baryon density& $0.0221\pm0.0002$\cr
$\Omega_{\rm c}h^2$& Cold dark matter density& $0.1187\pm0.0017$\cr
$\theta_\ast$& Acoustic angular scale& $0.010415\pm0.000006$\cr
$A_{\rm s}$& Amplitude of density perturbations&
 $(2.20\pm0.06)\times10^{-9}$\cr
$n$& Logarithmic slope of perturbations& $0.961\pm0.005$\cr
$\tau$& Optical depth due to reionisation& $0.092\pm0.013$\cr
\noalign{\vskip 3pt\hrule\vskip 4pt}
}}
\endtable
{
\par\noindent
{\bf Table~1}:\
6-parameter set describing the basic cosmology, derived from {\it Planck\/}
\cite{PlanckMission}
plus other data sets \cite{Bennett2012,ACT,SPT,Percival2010,Blake2011,Beutler2011,Padmanabhan2012,Anderson2013}.
See Ref.~\cite{PlanckParameters} for details.}\par
\bigskip

It is surprising that the Universe can be boiled down to just half a dozen
numbers, given the huge amount of cosmological information available
from the CMB, as well as from galaxy surveys and other astrophysical probes.
This dramatic compression of information
requires that the distribution of temperature anisotropies has
close to Gaussian statistics \cite{PlanckNG} in order for maps to be fully
described by power spectra.  In addition, the simplicity of the underlying
physics \cite{Dodelson} leads to the power spectra demonstrating a vastly
reduced number of degrees of freedom
compared with what one could imagine.  In a way, the large-scale and early
Universe is quite simple, being essentially uniform, with small amplitude
perturbations that are maximally random, i.e.\ with no correlated phases.
This means that the early perturbations have none of the
non-linear complexity required to describe today's small-scale objects
such as galaxies, planets and people.

Since the Universe is uncomplicated enough (at least in a statistical sense) to
be encapsulated in a few numerical factoids, then the simplest such quantities
should be much more familiar.
Every educated human should know some of the numbers that
describe their Cosmos, at least as well as the names of the local Solar
System planets,
and other facts, such as the dates of famous historical
events, or the statistics of a favourite sports team.

Innumerable quantities could be used to articulate our
present understanding of the Universe, and different cosmologists
have their own favourites.  Here, we select a few derived cosmic numbers,
and explain how modern precision cosmology affects different ways of
characterizing them.

Several quantities are easier for the
non-expert to grasp, compared to the standard set.
Others involve exploiting particular numerological coincidences -- but
we do not claim any special significance to those numbers we choose to
highlight.  Nevertheless, we hope that
some of these quantities may help you remember
your cosmic serial numbers, and grasp more fully the extent of our
present understanding of the Universe in which we live.

\section*{Cosmological data}

We use data constraints provided by the {\it Planck\/} satellite
\cite{PlanckMission}, which maps
the pattern of temperature variations on the microwave sky.  Such
CMB experiments probe the structure of the Universe at the time when
photons last interacted with matter significantly, the so-called
`last-scattering surface' about 370{,}000 years after the Big Bang.
The power spectrum (or equivalently the correlation function) of these
variations encodes information about the initial nature of the density
perturbations and how they have evolved over cosmic times.  Hence by
measuring them accurately, we can derive the parameters that describe the
large-scale Universe.  Previous CMB measurements, including from the
Wilkinson Microwave Anisotropy Probe ({\it WMAP\/}) satellite \cite{WMAP9},
showed that a fairly simple model, the SMC (also called `$\Lambda$ cold dark
matter' or $\Lambda$CDM), fits the data and requires
just six free parameters.  {\it Planck\/} has confirmed with
greater precision that this basic model still works well.

Table~1 lists the set of six parameters most directly measurable from the CMB.
The 6-parameter model requires a fixed framework, including a
set of testable assumptions (presented in Table~2).

\medskip
\nointerlineskip
\setbox\tablebox=\vbox{
\newdimen\digitwidth 
\setbox0=\hbox{\rm 0} 
\digitwidth=\wd0 
\catcode`*=\active 
\def*{\kern\digitwidth} 
\newdimen\signwidth 
\setbox0=\hbox{{\rm +}} 
\signwidth=\wd0 
\catcode`!=\active 
\def!{\kern\signwidth} 
\newdimen\pointwidth 
\setbox0=\hbox{\rm .} 
\pointwidth=\wd0 
\catcode`?=\active 
\def?{\kern\pointwidth} 
\halign{\tabskip=0.0em\DB{#}\hfil\tabskip=1.0em&
\DR{#}\hfil\tabskip=0.0em\cr
\noalign{\doubleline}
\multispan2\hfil Assumptions underlying the SMC\hfil\cr
\noalign{\vskip 3pt\hrule\vskip 4pt}
1& Physics is the same throughout the observable Universe.\cr
2& General Relativity is an adequate description of gravity.\cr
3& On large scales the Universe is statistically the same everywhere.\cr
4& The Universe was once much hotter and denser and has been expanding.\cr
5& There are five basic cosmological constituents:\cr
\omit& \DB{5a}\quad Dark energy behaves just like the energy density of the
 vacuum.\cr
\omit& \DB{5b}\quad Dark matter is pressureless (for the purposes of forming
 structure).\cr
\omit& \DB{5c}\quad Regular atomic matter behaves just like it does on Earth.\cr
\omit& \DB{5d}\quad Photons from the CMB permeate all of space.\cr
\omit& \DB{5e}\quad Neutrinos are effectively massless (again for structure
 formation).\cr
6& The overall curvature of space is flat.\cr
7& Variations in density were laid down everywhere at early times,\cr
\omit& \omit\hfil\DR{proportionally in all constituents.}\cr
\noalign{\vskip 3pt\hrule\vskip 4pt}
}}
\endtable
{
\par\noindent
{\bf Table~2}:\
Basic assumptions for the `Standard Model of Cosmology'.  Note that all of
these are testable, and have successfully passed the tests to date.  Because
of the dominance of dark matter (which is mostly `cold', CDM) and dark energy
(usually identified with the cosmological constant, $\Lambda$), the SMC is
often referred to as the `$\Lambda$CDM' model.}\par
\bigskip

There are many more things to measure about the Universe than the CMB, but
it provides a high-fidelity and well-understood data set that is
very powerful in combination with other kinds of data.
Following the Planck Collaboration we elect to use the constraints coming
from the {\it Planck\/} data combined with: large-angle
polarisation measurements from {\it WMAP\/} \cite{Bennett2012};
small scale (i.e.\ high multipole $\ell$) CMB data from the ACT \cite{ACT}
and SPT \cite{SPT} experiments; and a set of estimates of
the so-called `baryon acoustic oscillations'
\cite{Percival2010,Blake2011,Beutler2011,Padmanabhan2012,Anderson2013}
in the relatively nearby Universe.
Together, this data combination is described by the labels
{\it Planck\/}+WP+HighL+BAO and gives highly precise
and self-consistent determinations for the cosmological parameters.  Other
combinations of data could be chosen, which would make only slight
differences in the numerical values (some examples are shown in Fig.~1).

Reference~\cite{PlanckParameters}
describes how a Monte Carlo Markov chain approach
is used to fit cosmological models to the data, and hence to extract parameter
values and uncertainties.  These publicly available chains
allow calculation of probability distributions for any derived quantity,
and the determination of the most likely values and uncertainties;
these Markov chains are provided through the Planck Legacy Archive
\footnote{See {http://www.sciops.esa.int/index.php?project=planck\&page=Planck\_Legacy\_Archive}}.
From the full distributions for a derived quantity, we present the mean and
standard deviation ($\sigma$).  Since most parameters are detected with high
significance, the distributions are fairly bell-shaped (see Fig.~1),
indicating a reasonable characterisation of the constraints.

\begin{figure}[!t]
\centerline{\includegraphics[width=0.9\columnwidth]{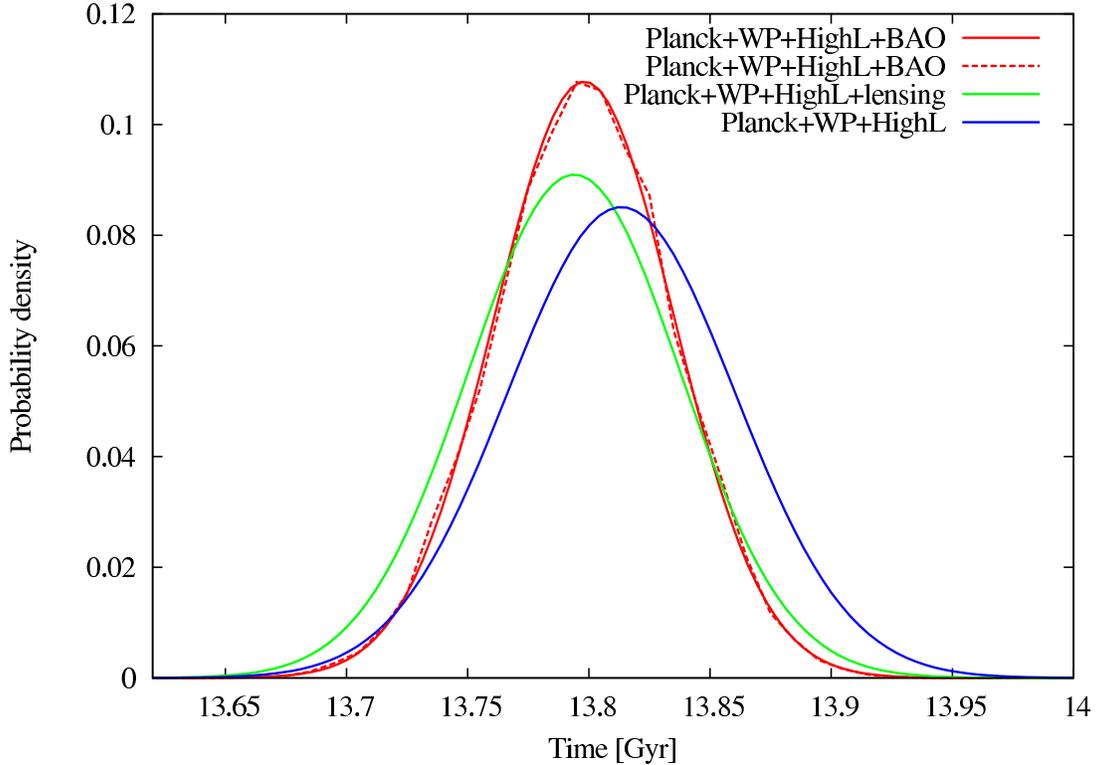}}
\caption{
Example of probability distributions, here for the
age of the Universe.  The dashed red line
shows the results directly obtained from the {\it Planck\/} chains, which is
well described by
a Gaussian distribution, as indicated by the solid red curve.  This plot is
specifically for the data combination coming from the 2013 release from
{\it Planck}, together with large-angle polarisation data from {\it WMAP\/}
(`WP'), additional constraints at large multipoles from SPT and ACT
(`HighL') and constraints on the scale of the acoustic oscillations in the
baryons at relatively low redshift (`BAO').  The other two Gaussians
show how different data combinations can give somewhat different (although
still statistically consistent) results.}
\end{figure}
\bigskip

\section*{Cosmological Quantities}

We will now discuss each derived quantity in turn.  The precision may be of
greater interest in some cases than others, and so we use the
following notation:
`=' means `essentially identical', i.e. within
about $1\,\sigma$ (note that this is empirical equivalence, and not the same
thing as mathematical equivalence);
`$\simeq$' means `pretty close', i.e.\ within $3\,\sigma$ or so;
and `$\sim$' means `roughly', i.e.\ similar in magnitude, but not
necessarily within a few $\sigma$.

\paragraph*{Age of the Universe}
Probably the conceptually simplest quantity is
the age of the Universe, $t_0$.  In the usual units
$t_0=(13.80\pm0.04)\,$Gyr, corresponding to $0.435$ exaseconds (in
S.I. units) or
$\simeq5$ trillion days.  Using the fine-structure constant
($\alpha\equiv e^2/4\pi\epsilon_0\hbar c \simeq1/137$,
a dimensionless number that gives
the strength of electromagnetism) then $t_0 \simeq 10^8/\alpha$ years.

The Earth and the rest of the Solar System formed approximately
$4.6\,$Gyr ago, e.g.\ Ref.~\cite{BouvierWadhwa} gives a precise age of
$(4.5682\pm0.0003)\,$Gyr.
This is essentially $t_0/3$ ago, so that Earth formed when the
Universe was 2/3 of its present age.

\paragraph{Other ways of telling the time}
An important parameter that describes our cosmological location is the
epoch, within the evolving model, at which we
are making our observations.  This epoch can be defined in
different ways.  The obvious way is to give the value of
$t_0$.  However, we can equivalently give the value of any of the time-evolving
parameters, for example the temperature of the CMB today,
which is $T_0=(2.7255\pm0.0006)\,$K \cite{Fixsen}.

Imagine a hypothetical situation in which we are communicating with
another `universe' where the physical constants might be
different -- then we would need to describe the epoch in dimensionless units
\cite{VariableG}.
For example, the CMB temperature
can be expressed dimensionlessly as a fraction of the electron
mass, $\Theta=kT_0/m_{\rm e}c^2\simeq4.6\times10^{-10} \simeq2^{-31}
 \simeq \alpha^4/(2\pi)$,
or $2.5\times10^{-13} \sim e^{-29}$ in terms of the proton mass.

We can also give our cosmic observational time by quoting the value of some
parameters at a fiducial epoch.  For example, the period when the matter and
radiation density had the same value, called `matter-radiation equality',
corresponds to redshift $z_{\rm eq}=3410\pm40$ (and this would be 1.69 times
higher if we compared matter with photons only).  This means that length scales
at the equality epoch were about 3400 times smaller than they are today in
the expanding Universe,
and the CMB temperature was then $9300\,$K, as hot as an A-type star.
The age at that epoch was $t_{\rm eq}=(51100\pm1200)\,$years.
And at that epoch the Universe was expanding much faster than today,
actually $H_{\rm eq}=(10.6\pm0.2)\,{\rm km}\,{\rm s}^{-1}\,{\rm pc}^{-1}$
(note this is per `pc', not `Mpc').

Another special epoch is when the CMB photons last significantly interacted
with matter, which is usually referred to as the epoch of `last-scattering'.
This corresponds to a redshift of $z_{\rm ls} =1089.3\pm0.4$
and a time of
$t_{\rm ls}=(372.8\pm1.5)\,$kyr.  At that epoch the CMB temperature was
close to $3000\,$K, the surface temperature of an M-type red dwarf star.

Alternatively, the formation of the Earth occurred
at a time corresponding to redshift $z_\oplus=0.420\pm0.005$, when the CMB
temperature was $(3.869\pm0.013)\,$K.  For an observer present when
Earth formed, today's epoch would be in the far future, and would correspond
to $z=-0.30\pm0.04$.

\paragraph*{Expansion rate}
In the expanding Universe, the `scale factor', $a(t)$, describes how length
scales evolve with time.  The derivative of this function evaluated today is
known as the Hubble constant, i.e.\ $H_0\equiv({\dot a}/a)|_{t_0}$ (where a
dot denotes a time derivative).  Since it measures the recession speed per
unit distance, the value of $H_0$ is usually given in units
of ${\rm km}\,{\rm s}^{-1}\,{\rm Mpc}^{-1}$, which is dimensionally the same
as the reciprocal of a time.  The value is
$H_0=(67.8\pm0.8)\,{\rm km}\,{\rm s}^{-1}\,{\rm Mpc}^{-1}$ or
$(2.20\pm0.02)\times10^{-18}\,{\rm s}^{-1}$ in inverse time units.
In fact $H_0\sim 1/t_0$, and observations were
precisely consistent with that value several years ago \cite{Frolop}.

However, the current value is slightly less than unity, $H_0 t_0=0.957\pm0.009$.
Since $H_0 t_0<1$ today, but $H$ tends to a finite value in the future while
$t$ increases without limit, then there must be a time in the future when
$H t=1$ exactly.  Let us call this special epoch the `Milne time', $t_{\rm M}$
(since in the empty universe proposed by E.A.\ Milne \cite{Milne}
$t$ is always $1/H$).  It will occur $(1.1\pm0.2)\,$Gyr from now,
i.e.\ when the Universe is about $15\,$Gyr old.

At the formation time of the Earth, corresponding to
$z_\oplus=0.42$, the Hubble parameter was
$(85.0\pm0.7)\,{\rm km}\,{\rm s}^{-1}\,{\rm Mpc}^{-1}$.
In the SMC, the Hubble parameter will approach
a value of $H_\infty\equiv H_0\times\Omega_\Lambda^{1/2}$ in the far
future.  This quantity is independent of the observer epoch and hence is,
in some sense, more fundamental than the Hubble `constant' today.
Its value is $H_\infty=(56.4\pm1.1)\,{\rm km}\,{\rm s}^{-1}\,{\rm Mpc}^{-1}$
$=\sqrt{10/3}\,$attohertz.
This means that $1/H_\infty=(17.3\pm0.3)\,$Gyr and
$H_\infty t_0=0.796\pm0.013$.

\paragraph*{Deceleration, jerk, snap, crackle and pop}
The Hubble constant is the slope of the scale factor $a(t)$ today, specifically
$H\equiv ({\dot a}/a)|_{t=t_0}$.  Dimensionless parameters can be defined to
describe higher-order derivatives of $a(t)$, namely:
deceleration, $q_0\equiv -(a {\ddot a}/{\dot a}^2)_{t=t_0}$;
jerk, $j_0\equiv (a^2 {\dddot a}/{\dot a}^3)_{t=t_0}$;
snap, $s_0\equiv (a^3 {\ddddot a}/{\dot a}^4)_{t=t_0}$;
crackle, $c_0\equiv (a^4 {\dddddot a}/{\dot a}^5)_{t=t_0}$
and pop, $p_0\equiv (a^5 {\ddddddot a}/{\dot a}^6)_{t=t_0}$
\cite{Visser2004}.
Fitting these quantities (now using a model that includes curvature as a
free parameter), we find:
$q_0=-0.537\pm0.016$;
$j_0=1.000\pm0.003$;
$s_0=-0.39\pm0.05$;
$c_0=3.22\pm0.12$;
and $p_0=-11.5\pm0.7$.

The dominance of matter makes the Universe decelerate at early times, and
dark energy drives the more recent accelerated expansion.  The cross-over
occurred when the deceleration was equal to zero, i.e.\ $q=0$, which
occurred at $z_q=0.649\pm0.027$.  This is somewhat earlier than the epoch when
$\Omega_{\rm m}=\Omega_\Lambda$, which occurred at $z_\Lambda=0.31\pm0.02$
(and note that those epochs cannot be coincident if the dark energy
behaves exactly like a cosmological constant).
It may be interesting to note that the formation of the Earth
(at $z=0.42$) is bracketed by these two epochs, specifically about
a billion years before dark energy dominated the cosmological energy budget, a
and a billion and half years after the Universe started to accelerate.

\paragraph*{Constituents}
The census of the contents of the Universe is
usually described in terms of the contribution to the average energy density,
as a fraction of $\rho_{\rm crit}$, which is the critical value that makes
space curvature flat.  So we have
$\Omega_{\rm b}\equiv\rho_{\rm b}/\rho_{\rm crit}$ ($=0.0482\pm0.0009$)
for the baryon abundance,
$\Omega_{\rm c}\equiv\rho_{\rm c}/\rho_{\rm crit}$ ($0.260\pm0.010$)
for (cold) dark matter,
the total for matter being $\Omega_{\rm m}\equiv \Omega_{\rm b}+\Omega_{\rm c}$
($=0.308\pm0.010$)
and $\Omega_\Lambda\equiv\rho_\Lambda/\rho_{\rm crit}$
($=0.692\pm0.010$) for the cosmological
constant or `dark energy'.  Based on the current estimate for $H_0$, we find
$\rho_{\rm crit}\equiv 3H_0^2/8\pi G
=(8.6\pm0.2)\times10^{-27}\,{\rm kg}\,{\rm m}^{-3}$.  This value is equivalent
to the mass of about 5 protons or neutrons per cubic
metre of space (imagine an atom of mass number 5 in each m$^3$
-- easy to remember, since there {\it are\/} no nuclei of mass number 5 which
are even remotely stable).  In contrast, the abundance of baryons corresponds
to approximately one in every sphere of $1\,$m radius.

The cosmological constant is usually written as $\Lambda$ and is the same
quantity that appears in Einstein's field equations.  It has units of
inverse seconds squared and is related to the equivalent mass density in this
component through $\Lambda=8\pi G \rho_\Lambda$.  The data give
$\Lambda=(1.00\pm0.04)\times10^{-35}\,{\rm s}^{-2}$.
It can be expressed in SI units using only three words:
`ten square attohertz'.  It can also be written as
$\Lambda=(10.0\,{\rm Gyr})^{-2}$.
In everyday units one can express the equivalent
vacuum mass density as
$\rho_\Lambda=(6.0\pm0.2)\times10^{-27}\,{\rm kg}\,{\rm m}^{-3}$.  Since
$p=-\rho c^2$ for vacuum energy, then the pressure is $-5.4\times10^{-10}\,$Pa
or $-4.0\times10^{-12}\,$Torr, or $-5.3\times10^{-15}$ atmospheres.

The values of $\Omega_{\rm b}h^2$ and $\Omega_{\rm c}h^2$ are conventional
parameters, given in Table~1.  In S.I.\ units we have
$\rho_{\rm b}=(4.16 \pm 0.05)\times10^{-28}\,{\rm kg}\,{\rm m}^{-3}$,
$\rho_{\rm c}=(2.23 \pm 0.03)\times10^{-27}\,{\rm kg}\,{\rm m}^{-3}$
and $\rho_{\rm m}=(2.65 \pm 0.04)\times10^{-27}\,{\rm kg}\,{\rm m}^{-3}$.
One can easily define ratios of the $\Omega$s, e.g.\
$\Omega_\Lambda/\Omega_{\rm m}=2.25\pm0.11$ and
$\Omega_{\rm m}/\Omega_{\rm b}=6.39\pm0.11$.
It may be interesting to note that
$\Omega_{\rm c}/\Omega_{\rm b} = 2\Omega_\Lambda/\Omega_{\rm c}$
($=5.36$).

For the relativistic particle content
$\Omega_{\rm r} = (9.0\pm0.2)\times10^{-5}$ today
(including 3 species of massless neutrinos), or
$\Omega_\gamma=(5.38\pm0.12)\times10^{-5}$ = $\alpha^2$ (for photons only).

The baryon-to-photon ratio, defined conventionally through
$n_{\rm b}/n_\gamma\equiv \eta \equiv \eta_{10}\times10^{-10}$
is given by $\eta_{10}=6.13\pm0.08\simeq 2\pi$ (with helium abundance being
a free parameter in this particular calculation).

\paragraph*{Initial conditions}
So far, all the quantities describe a perfectly smooth Universe.  However,
we know there are imperfections in this picture, density irregularities
laid down at early times that grew through
gravitational instability into the rich structure seen today.
There are several ways to parameterise the amplitude of the initial
perturbations, with the conventional way being through the amplitude
of the power spectrum of the Fourier modes.  For example, the {\it Planck\/}
team give $A_{\rm s}=(22.0\pm0.6)\times10^{-10}$
(actually they use $\log A_{\rm s}$)
at wavenumber $k=0.05\,{\rm Mpc}^{-1}$.

As an alternative, one can consider the `lumpiness' of the density field
directly.  This is often expressed as the standard deviation of the
variations in density, i.e.\ the square root of the variance $\sigma_R^2$ of
$\delta\rho/\rho$, in spheres of a given radius, $R$.  A conventional choice is
to use a radius of $8\,h^{-1}\,{\rm Mpc}$; this gives
$\sigma_8 =0.826\pm0.012$, where the $h^{-1}$ scaling is a remnant from a
time when the Hubble constant was very poorly known.  Instead of using
the somewhat obscure $\sigma_8$ parameter, one could instead ask for the
size of sphere for which the variance is precisely unity -- this turns out
to be $R_{\sigma{=}1}=(8.9\pm0.3)\,$Mpc
(and note the lack of $h$ scaling here).

Another way to define the amplitude would be to take the value of the
density perturbation at the Hubble scale (defined explicitly through $k=aH$)
at some special epoch, say the Milne time $t_{\rm M}$.
This gives $\sigma_{\rm M}=(5.6\pm0.3)\times10^{-6}$ ($=3^{-11}$),
which could be considered a more
observer-independent measure of the fluctuation amplitude.

In the simplest pictures for these density perturbations, they would
be laid down in a way that is democratic with respect to scale -- the
so-called Harrison-Zeldovich initial conditions.  This corresponds to a
logarithmic variation of power with scale, denoted by `$n$'
(i.e.\ $n\equiv d\ln{P(k)}/d\ln{k}$) with the default value being unity.
In fact, there seems to be a little more power on large scales compared to
small scales, such that $n=0.961\pm0.005$.  This is seen by many cosmologists
as support for an idea like cosmic inflation for the origin of the
perturbations.

It may be interesting to
note the coincidence that $n= H_0 t_0$.
In fact $n/(H_0t_0)=1.004\pm0.007$.

Another way to describe perturbations focuses on how they are growing
today.  In the $\Lambda$CDM model, this is strongly affected by the presence of
a cosmological constant, which impedes the amplification of structure at
relatively recent times.  Relative to a flat model with vanishing $\Lambda$,
the `growth suppression factor' is $g=0.784\pm0.006$.  

\paragraph*{Curvature}
Although we do not know if the whole extent of space is finite or infinite,
we can measure curvature within our Hubble patch.  
{\it Planck\/} (together with other data sets, see \cite{PlanckParameters})
yields
$\Omega_{K}=-0.000\pm0.003)$, where $\Omega_{K}=1-\Omega_{\rm tot}$.
This means that the total density (in matter plus radiation plus dark energy)
is quite accurately given by $\rho_{\rm crit}$.

Constraints can be placed on the radius of curvature, such that
$R_{\rm curv}/R_{\rm H} > 12$ (at 95\% confidence,
with $R_{\rm H}\equiv c/H_0$).  The particle horizon is also well-defined,
$R_{\rm p}= X c/H_0$, with $X=3.21\pm0.04$.
For the distance to the last-scattering
surface (before which the Universe is optically thick to CMB photons), we
find $X= 3.15\pm0.04$.
Using this to define an observable volume and considering constraints on
curvature, we can derive a lower limit to the number of such volumes in
the {\it entire\/} Universe
(assuming that our own patch is a fair sample of course):
$N_{\rm U}>250$ \cite{ScottZibin}.

\paragraph*{Observable Universe}
We cannot say whether there are an infinite number of particles in the
entire Universe.  However, we can determine the number in the
{\it observable\/} Universe, which has a finite volume.
Using the above definition of the
observable distance (as the distance to the last-scattering surface),
and assuming flat geometry, we find that the radius of the
observable Universe is $(429.2\pm1.3)\,$Ym (with the particle horizon being
only about 2\% larger, $(437.9\pm1.3)$Ym).
Here the prefix `Y' is for
`Yotta', meaning $10^{24}$, the largest approved SI unit multiplier.
It may be a coincidence that, for the sizes of {\it anything\/} observable
in metres, we do not need a larger prefix.

The total number of baryons contained within the observable Universe is then
$N_{\rm b}=(8.27\pm0.11)\times 10^{79}$.  For photons we have
$N_\gamma=(1.360\pm0.012)\times10^{89}$, and the total
number of known particles (dominated by photons and massless neutrinos) is
$N_{\gamma+\nu}=(2.49\pm0.02)\times10^{89}$ ($\sim \alpha^{-42}$).

\paragraph*{Acoustic scales}
The CMB variations are largely determined by oscillating sound waves, with a
wide range of wavelengths.  Because of the finite speed of propagation of
these acoustic modes, and the finite age of the Universe, a
characteristic scale is built in by the physics.  At the distance of the
last-scattering surface this length scale projects onto a
particular angular scale, which is effectively the angular size
of `blobs' in CMB maps.  In conventional units, this scale is
$\theta_\ast=0.5968^\circ\pm0.0003^\circ \simeq0.6^\circ$.  This is
essentially the same as (only about 10\% larger than) the angular diameter of
the Sun and the Moon.

\paragraph*{Rescattering}
A fraction of the CMB photons are scattered in a period of relatively recent
reionisation of the Universe.  This is often expressed as an optical depth,
but more directly, the rescattered fraction is about 8.8\%.  The distance
out to which the Universe is ionised, i.e.\ the distance to the reionisation
surface, is $(305\pm6)$Ym.

\paragraph*{\bf Planck units}
The quantities that describe the Universe could be given in
different systems of units.  The system of `Planck units' is formed by
using the speed of light ($c$), reduced Planck constant ($\hbar$), and
gravitational constant ($G$) to form the
Planck length ($l_{\rm P}=\sqrt{\hbar G/c^3}$),
Planck time ($t_{\rm P}=\sqrt{\hbar G/c^5}$),
Planck mass ($m_{\rm P}=\sqrt{\hbar c/G}$), and
Planck temperature ($T_{\rm P}=\sqrt{\hbar c^5/G k^2}$).
In these units, we have:
$t_0=(8.08\pm0.02)\times10^{60}\,t_{\rm P}$
 ($\simeq 5\times2^{200}\,t_{\rm P}$);
$H_0=(1.185\pm0.013)\times10^{-61}\,t_{\rm P}^{-1}$;
$\Lambda=(2.91\pm0.12)\times10^{-122}\,t_{\rm P}^{-2}$; and the
CMB temperature today $T_0\simeq(1/41) 2^{-100}\,T_{\rm P}$,
or $T_0/T_{\rm P}=(160/3^8)2^{-100}$.  So to use an analogy with the
musical scale, one can say that the CMB temperature today is one hundred
octaves, eight perfect fifths, and one justly tuned minor fifth below the
Planck temperature.

The particle content of the Universe is related to the total entropy.
One can define the asymptotic `Gibbons-Hawking entropy' \cite{GibbonsHawking}
for de Sitter
space as 1/4 the asymptotic cosmological horizon area in Planck units, i.e.\
$S/k\equiv 3\pi/(\Lambda t_{\rm P}^2)\simeq 5(t_0/t_{\rm P})^2$.  This is
$(3.24\pm0.12)\times10^{123} \simeq 5^3 2^{400}$.

\section*{Mnemonic cosmology}

Martin Rees wrote a popular cosmology book entitled `Just Six Numbers'
\cite{Rees}.
Although his numbers differ from the six which are well measured
in today's cosmological data, the basic message is the same: we have developed
an understanding of the large-scale Universe that is rather simple,
is describe by roughly a handful of numbers, and if they had other values the
Universe would be quite different.  The
Standard Model of Cosmology is built on a framework of assumptions which
are reasonable and few in number.  Within that framework only half a dozen
parameters are required to fit the current data.  However, we have several
choices for how to present these numbers, including the epoch at which to
specify them, the units to use, and whether to focus on dimensionless
ratios.  Since these are the quantities which describe the entire cosmos,
then it is worth manipulating and evaluating them, in order to better grasp how
our Universe measures up.

Lots of different numbers have been presented here, with the expectation
that distinct choices might appeal to different people.
In Table~3 we have gathered
together some of our favourite numerical facts about the whole Universe.

\nointerlineskip
\setbox\tablebox=\vbox{
\newdimen\digitwidth 
\setbox0=\hbox{\rm 0} 
\digitwidth=\wd0 
\catcode`*=\active 
\def*{\kern\digitwidth} 
\newdimen\signwidth 
\setbox0=\hbox{{\rm +}} 
\signwidth=\wd0 
\catcode`!=\active 
\def!{\kern\signwidth} 
\newdimen\pointwidth 
\setbox0=\hbox{\rm .} 
\pointwidth=\wd0 
\catcode`?=\active 
\def?{\kern\pointwidth} 
\halign{\tabskip=0.0em\DB{#}\hfil\tabskip=1.0em&
\tabskip=1.0em\DR{#}\hfil\tabskip=1.0em&
\DG{#}\hfil\tabskip=0.0em\cr
\noalign{\doubleline}
\multispan3\hfil Cosmic Mnemonics\hfil\cr
\noalign{\vskip 3pt\hrule\vskip 4pt}
Symbol& Quantity& Value \cr
\noalign{\vskip 3pt\hrule\vskip 4pt}
$t_0$& Age of the Universe today& $\simeq 5\,$trillions days
 $\simeq5\times2^{200}t_{\rm P}$\cr
$\Lambda$& Cosmological constant& $= 10^{-35}\,{\rm s}^{-2}=$ ten square
 attohertz\cr
$H_0 t_0$& Expansion rate times age today& $\simeq0.96=n$\cr
$H_\infty$& Future limit for Hubble parameter&
 $\simeq56\,{\rm km}\,{\rm s}^{-1}\,{\rm Mpc}^{-1}
  =\sqrt{10/3}\,$attohertz\cr
$z_q$& Redshift at which acceleration was zero& $\sim0.65$\cr
$z_\oplus$& Redshift of formation of the Earth& $=0.42$\cr
$\theta_\ast$& Characteristic scale of CMB anisotropies&
 $\simeq0.6^\circ$ $\sim$ solar angular diameter\cr
$\Omega_\gamma$& Density parameter for photons& $=\alpha^2$\cr
$\eta_{10}$& Baryon-to-photon ratio ($\times10^{10}$)& $\simeq2\pi$\cr
$R_{\rm obs}$& Radius of the observable Universe& $\sim400\,$Ym\cr
$N_{\rm part}$& Number of particles in observable Universe&
 $={\rm few}\times10^{89}\sim\alpha^{-42}$\cr
$R_{\sigma{=}1}$& Scale for density contrast of unity& $\simeq9\,$Mpc\cr
$\sigma_{\rm M}$& Hubble-scale perturbation at Milne epoch&
 $\simeq6\times10^{-6}$\cr
\noalign{\vskip 3pt\hrule\vskip 4pt}
}}
\endtable
{
\par\noindent
{\bf Table~3}:\
A selection of numbers that describe our Universe.}\par
\bigskip

\vfil\eject

\section*{Supplementary Materials}

\paragraph*{Numerology in cosmology}
Cosmology has a long history of `numerology' \cite{Barrow}, with attempts to
connect apparent coincidences in order to motivate fundamental theories.
Many well-known scientists are connected with this topic,
including Eddington, Dirac, Teller, Dicke, and Weinberg.

The Standard Model of Particle Physics contains about 26 parameters, none of
which can be determined from first principles -- although most theorists
expect that they will one day emerge from a smaller set of parameters
in a more fundamental theory \cite{TARW}.
Cosmology brings in some additional numbers.
A discussion of where all these parameters come from is often couched in
terms of anthropic arguments, dealing with the Multiverse or the Landscape.

As a specific example, Martin Rees focuses on `Just Six Numbers' in his 1999
book, describing how the Universe could be utterly different if some
quantities had different values.  The set of numbers focussed on there is
different from those used in the Standard Cosmological Model.  Explicitly
Rees' numbers are: $N$ ($\simeq10^{36}$),
the ratio of the fine structure constant to the gravitational coupling
constant using protons;
$\epsilon$ ($\simeq0.007$), the fraction of mass released as energy when H
fuses to He;
$\Omega$ ($=\Omega_{\rm M}\simeq0.3$);
$\lambda$ ($=\Omega_\Lambda\simeq0.7$);
$Q$ ($\simeq10^{-5}$), the binding energy per rest mass ratio for
large-scale gravitationally bound objects;
and $D$ ($=3$), the number of macroscopic space dimensions.

\paragraph*{Details on parameter fits}
We base our numerical constraints on the Markov chains produced by the
Planck Collaboration, as described in detail in
Refs.~\cite{PlanckLikelihood,PlanckParameters}.  The CMB multipole power
spectra are estimated from foreground-substracted {\it Planck\/} maps
using a likelihood approach.  These experimentally determined power spectra
are compared with theoretical spectra
computed for a given set of parameters using the code
{\tt camb} \cite{CAMB}.  A set of base parameters (including the main
six cosmological ones, plus various others, such as calibration coefficients)
are searched through using the code {\tt CosmoMC} \cite{COSMOMC}.
The chains produced
by this code give the correct posterior distribution for the parameters,
including their correlations.  Hence one can easily extract the statistics
for a derived parameter by plotting a histogram for that parameter directly
from the chain.  From those distributions we simply extract the mean and
standard deviation, and give those as the central value and uncertainty.

When we fit for the variance of the density field $\sigma_R^2$ we are
implicitly doing this for the linear power spectrum, i.e.\ neglecting
non-linear effects, which are important for small scales at late times.
Hence the value of $\sigma_R$ derived is not {\it quite\/} the value one
would actually obtain by smoothing today's density field in spheres of
size $R$.

We have assumed throughout that the dark energy is precisely a cosmological
constant, i.e.\ the `equation of state' of the dark energy is given by
$w=-1$, where $w\equiv p_\Lambda/\rho_\Lambda c^2$.
There are several different criteria one could use for defining when dark
energy starts to dominate.  Two obvious examples are when $q=0$ and when
$\Omega_{\rm m}=\Omega_\Lambda$, which we define as the redshifts
$z_q$ and $z_\Lambda$, respectively.  For $w=-1$ these are necessarily
different, but they would coincide if $w=-1/2$.

When we fit for deceleration, jerk, snap and crackle, we use chains which
include the curvature as a free parameter.  The reason for this is that
otherwise some of the quantities have trivial values, e.g.\
$j_0=1+2\Omega_{\rm r}-\Omega_{K}$, which would be unity to four
significant figures without curvature being allowed to vary.

The last-scattering epoch could be defined in several different ways.  We
specifically use the redshift of the peak of the `visibility function', i.e.\
the function describing the probability of scattering per unit redshift
interval (ignoring the effect of reionization).
Other criteria, such as the epoch at
which half of the hydrogen was ionised, or where the Thomson optical depth is
unity, would give different numerical values.

When quoting numerical values we use the convention that an error bar requires
only a single digit (unless it is `1', in which case two digits are used), and
then the central value is quoted to the corresponding number of digits.

The precise central values of the parameters today will of course change
when improved observations become available, with deviations of 2 or even
$3\,\sigma$ being not unreasonable.  Hence some of the numerical
coincidences described here may not be quite accurate in future.

\paragraph*{Simplified cosmology}
We can present a simplified version of the overall cosmological
model, extending an earlier suggestion in Ref.~\cite{Page}.
In the spatially flat $\Lambda$CDM Friedmann-Lema{\^\i}tre-Robertson-Walker
model at late times, when radiation is not dynamically important,
we can write $a(t) = \{(2/3)\sinh[(3/2)H_\infty t]\}^{2/3}$.  Here the
coefficient has been chosen so that
$a\propto (H_\infty t)^{2/3}$ for $H_\infty t \ll 1$.
Then empirically $a_0 = 1$, within $1\,\sigma$.  This formula enables us to
convert rather straightforwardly between age and scale factor (or redshift).
Using the fact that
$\Lambda = 3 H_\infty^2$ is within $1\,\sigma$ of ten square attohertz, of
$(10\,{\rm Gyr})^{-2}$, and of $3\pi/(5^3 2^{400})$ in Planck units,
we can then derive the age
$t_0 = (2/\sqrt{3\Lambda})\ln{(1.5+\sqrt{3.25})}$, which gives $4.36\times
10^{17}$ s or 13.80 Gyr or $8.07\times 10^{60} t_{\rm P}$, and
1/$H_0 = 9/\sqrt{39\Lambda}$, which gives $4.56\times 10^{17}\,$s or
14.41\,Gyr (equivalent to
$H_0 = 67.85\,{\rm km}\,{\rm s}^{-1}\,{\rm Mpc}^{-1}$)
or $8.43\times 10^{60} t_{\rm P}$
in either seconds, gigayears, or Planck units.

We can also deduce $\Omega_\Lambda = 9/13 = 0.692$, $\Omega_{\rm m} = 4/13
= 0.308$, $q_0 = -7/13 = -0.538$, $j_0 = 1$, $s_0 = -5/13 = -0.385$, $c_0 =
541/169 = 3.20$, and $p_0 = -25073/2197 = -11.4$ (assuming $\Omega_{K} =
0$ and ignoring $\Omega_{\rm r} \sim 10^{-4}$). $\Omega_{\rm c}/\Omega_{\rm
b} = 2\Omega_\Lambda/\Omega_{\rm c}$ allows us further to deduce
$\Omega_{\rm b} = (13-3\sqrt{17})/13 = 0.0485$ and $\Omega_{\rm c} =
(3\sqrt{17}-9)/13 = 0.259$ (as empirical equalities, correct within
$1\,\sigma$).  Another mnemonic is that back when the temperature was much
higher than all the neutrino masses, the total radiation density was very
nearly $\rho_{\rm r} = (10/9) T_0^4 (a_0/a)^4$.  If this applied today,
using $T_0 = (160/3^8) 2^{-100}$ in Planck units would give $\Omega_{\rm r}
= 40^8/(3^{33} 13) = 9.07\times 10^{-5}$.  Similarly, one obtains
$\Omega_\gamma = 2^{23} 5^6 \pi^2/(3^{32} 13) = 5.37\times 10^{-5}$.

\bibliography{mnemonic}

\bibliographystyle{Science}

\begin{scilastnote}
\item This research was supported by the Natural Sciences and Engineering
Research Council of Canada and the Canadian Space Agency.
We thank Hilary Feldman for useful comments on
an earlier draft, and members of the Planck Collaboration, particularly Jim
Zibin, for helpful discussions.
\end{scilastnote}

\end{document}